\documentclass{article}

\usepackage[numbers,sort&compress]{natbib}
\usepackage[affil-it]{authblk}
\usepackage[utf8,latin1]{inputenc} 
\usepackage[scaled=0.90]{inconsolata}
\usepackage[T1]{fontenc}    
\usepackage[hidelinks]{hyperref}       
\usepackage{url}            
\usepackage{booktabs}       
\usepackage{amsfonts}       
\usepackage{nicefrac}       
\usepackage{microtype}      
\usepackage{xcolor}         
\usepackage{transparent}
\usepackage{multirow}
\usepackage{todonotes}
\usepackage{enumitem}
\usepackage{graphicx} 
\usepackage{amsmath,amssymb,amstext}
\usepackage{xspace}
\usepackage[version=3]{mhchem}
\usepackage{xr}
\makeatletter
\newcommand*{\addFileDependency}[1]{
\typeout{(#1)}
\@addtofilelist{#1}
\IfFileExists{#1}{}{\typeout{No file #1.}}
}\makeatother

\newcommand*{\myexternaldocument}[1]{%
\externaldocument{#1}%
\addFileDependency{#1.tex}%
\addFileDependency{#1.aux}%
}
\myexternaldocument{si}

\usepackage{glossaries}
\newacronym{oc20}{OC20}{Open Catalyst 2020}
\newacronym{oc22}{OC22}{Open Catalyst 2022}
\newacronym{ocp}{OCP}{Open Catalyst Project}
\newacronym{oer}{OER}{oxygen evolution reaction}
\newacronym{her}{HER}{hydrogen evolution reaction}
\newacronym{orr}{ORR}{oxygen reduction reaction}
\newacronym{dft}{DFT}{density functional theory}
\newacronym{bfgs}{BFGS}{Broyden-Fletcher-Goldfarb-Shanno}
\newacronym{cg}{CG}{conjugate gradient}
\newacronym{ml}{ML}{machine learning}
\newacronym{mlp}{MLP}{machine learning potential}
\newacronym{gpmin}{GPMin}{Gaussian process minimizer}
\newacronym{gp}{GP}{Gaussian process}
\newacronym{ase}{ASE}{the Atomic Simulation Environment Python package}
\newacronym{gnn}{GNN}{Graph neural network}
\newacronym{pca}{PCA}{principal component analysis}
\newacronym{ewt}{EwT}{Energies within Threshold}
\newacronym{vasp}{VASP}{the Vienna Ab initio Simulation Package}
\newacronym{mae}{MAE}{mean absolute error}
\newacronym{scf}{SCF}{self-consistent field}
\newacronym{rbf}{RBF}{radial basis function}
\newacronym{cbf}{CBF}{circular basis function}
\newacronym{sbf}{SBF}{spherical basis function}
\newacronym{pbe}{PBE}{Perdew-Burke-Ernzerhof}
\newacronym{rpbe}{RPBE}{revised Perdew-Burke-Ernzerhof}
\newacronym{vdw}{vdW}{van der Waals}
\newacronym{beefvdw}{BEEF-vdW}{Bayesian error estimation functional with van der Waals correlation}
\newacronym{abi}{VASP-BFGS}{VASP Interactive with the BFGS optimizer}
\newacronym{mof}{MOF}{metal-organic-framework}
\newacronym{ipa}{IPA}{isopropyl alcohol}
\newacronym{rmsd}{RMSD}{root-mean-square deviation}

\title{Generalization of Graph-Based Active Learning Relaxation Strategies Across Materials}
\author[1, *]{Xiaoxiao Wang}
\author[1, *]{Joseph Musielewicz}
\author[1]{Richard Tran}
\author[2]{Sudheesh Kumar Ethirajan}
\author[3]{Xiaoyan Fu}
\author[1]{Hilda Mera}
\author[1, $\dagger$]{John R. Kitchin}
\author[4, $\dagger$]{Rachel C. Kurchin}
\author[1, 5, $\dagger$]{Zachary W. Ulissi}

\affil[1]{Department of Chemical Engineering, Carnegie Mellon University}
\affil[2]{Department of Chemical Engineering, University of California, Davis}
\affil[3]{Dalian Institute of Chemical Physics, Chinese Academy of Sciences}
\affil[4]{Department of Materials Science and Engineering, Carnegie Mellon University}
\affil[5]{Fundamental Artificial Intelligence Research, Meta Platforms, Inc., Menlo Park, CA}

\affil[*]{These authors contributed equally to this work}

\affil[$\dagger$]{Corresponding authors: jkitchin@andrew.cmu.edu, rkurchin@cmu.edu, zulissi@meta.com}

\date{October 2023}

\begin{document}

\maketitle


\begin{abstract}
Although density functional theory (DFT) has aided in accelerating the discovery of new materials, such calculations are computationally expensive, especially for high-throughput efforts. This has prompted an explosion in exploration of machine learning assisted techniques to improve the computational efficiency of DFT. In this study, we present a comprehensive investigation of the broader application of Finetuna, an active learning framework to accelerate structural relaxation in DFT with prior information from Open Catalyst Project pretrained graph neural networks. We explore the challenges associated with out-of-domain systems: alcohol ($C_{>2}$) on metal surfaces as larger adsorbates, metal-oxides with spin polarization, and three-dimensional (3D) structures like zeolites and metal-organic-frameworks. By pre-training machine learning models on large datasets and fine-tuning the model along the simulation, we demonstrate the framework's ability to conduct relaxations with fewer DFT calculations. Depending on the similarity of the test systems to the training systems, a more conservative querying strategy is applied. Our best-performing Finetuna strategy reduces the number of DFT single-point calculations by 80\% for alcohols and 3D structures, and 42\% for oxide systems.

\end{abstract}

\section{Introduction}
\label{sec:intro}
The urgent need to address climate and societal challenges has prompted the search for innovative solutions, and new material design and discovery have emerged as critical pathways. For instance, new catalysts are being explored for various crucial chemical reactions such as nitrogen reduction to replace the energy-intensive Haber-Bosch process \cite{Li2021, Foster2018}, and alcohol dehydrogenation for energy storage applications \cite{Bonitatibus2015}. Metal-oxides are being investigated as catalysts for oxygen reduction and evolution reactions in fuel cells \cite{Ge2015, Beall2021}. With the continued advancements in simulation tools and increased computing power, we can harness the power of computational chemistry to further accelerate the process of materials discovery. By utilizing computational techniques, researchers can rapidly screen a vast space of chemical compositions, structures, and properties that would otherwise be prohibitively expensive and time-consuming to explore experimentally \cite{Petousis2017, Lu2021}. Simulations can also provide valuable insights into the underlying reaction mechanisms, allowing for the optimization of material performance.

However, the computational cost of first-principles simulations becomes infeasible for large chemical spaces. In these cases, one prominent approach in accelerating material discovery is through the use of \gls{ml} methods, which leverage large databases and advanced algorithms to predict the properties and performance of materials. A properly-trained ML model is vastly lower in computational cost than computational tools such as \gls{dft} and can have comparable accuracy, provided it is being applied on sufficiently similar systems to the training data \cite{Schmidt2019}. The development of large material datasets, such as the \gls{oc20} Dataset and \gls{oc22} Dataset for metal and metal-oxide materials \cite{oc20,oc22}, the CoRE \cite{Chung2019}, CSD MOF \cite{Moghadam2017}, QMOF \cite{Rosen2021}, MOFX-DB \cite{Bobbitt2023} for \gls{mof}s, and Database of Zeolite Structures \cite{IZA-SC} and Zeo-1 \cite{Komissarov2022} for zeolites, has enabled data-driven materials modeling research on these materials classes. Remarkable progress has been made in the development of \gls{ml} models for accurate calculation of interatomic potentials and prediction of materials properties. \gls{gnn} models are specifically designed to handle the inherent graph-like structure of molecules and crystals, where atoms and bonds are represented as nodes and edges in a graph. This allows \gls{gnn} models to capture the complex interactions between atoms. \gls{gnn} models such as GemNet \cite{gemnet}, GemNet-OC \cite{gemnetoc}, SCN \cite{Zitnick2022}, eSCN \cite{Passaro2023}, and M3GNeT \cite{Chen2022}, can learn vast amounts of chemical information from training data, facilitating the efficient screening of materials and identifying promising candidates for specific applications. For example, the present leading \gls{ml} model on the \gls{ocp} Leaderboard can achieve a total energy \gls{mae} as low as 0.22 eV on \gls{oc20}-like catalyst structures.

Training \gls{ml} models on large datasets can be time-consuming and computationally demanding, especially in the field of molecular modeling, where diverse chemical structures and properties are involved. One approach to mitigate this challenge is to utilize transfer learning. Transfer learning allows a model to leverage the knowledge and representations learned during pretraining to initialize its parameters for a new dataset, significantly reducing the need for extensive data and computational resources in training \cite{Kolluru2022}. In our prior work, we proposed the Fine-Tuning Accelerated molecular simulations framework (Finetuna) \cite{finetuna} as a promising implementation of online transfer learning with pre-trained \gls{ml} models. Finetuna utilizes an active querying strategy to determine when to perform a \gls{dft} calculation and fine-tune the \gls{ml} model using the \gls{dft} results. More specifically, we benchmarked Finetuna performance on a set of catalyst systems from the \gls{oc20} validation set with a GemNet model trained on the \gls{oc20} training data. Finetuna demonstrated that local optimizations of \gls{oc20}-like systems can be greatly accelerated without sacrificing accuracy, as evidenced by a reduction of 90\% \gls{dft} calculations compared to a baseline approach. We note that the Finetuna framework is one of many online active learning approaches for accelerating atomistic simulations.
Other on-the-fly active learning frameworks typically start from scratch, using Gaussian process models or simpler neural network potentials \cite{Tran2018, Zhong2020, Vandermause2020, Vandermause2021, Yang2021, Shuaibi2020}.

In this work, we seek to extend Finetuna to different out-of-domain chemical systems and gain insights into the capabilities and limitations of the workflow. We conduct a case study of the Finetuna workflow on three groups of materials, namely ($C_{>2}$) alcohols, metal-oxides, and three-dimensional (3D) structures, for different applications. While the \gls{oc20} dataset is limited to the exploration of small ($C_{\leq2}$) adsorbates, we explore the adsorption of $C_{>2}$  alcohol-to-ketone intermediates in this work. We also explore metal-oxide catalysts and the structures of zeolites and \gls{mof}s which significantly deviate from the catalyst surfaces in the original \gls{oc20} dataset. Finally we examine the effect of spin polarization and systems with significantly more atoms than those in the training dataset, both factors that significantly decrease computational speed of \gls{dft}. We introduce new querying strategies for these systems with our best-performing Finetuna strategy capable of reducing \gls{dft} calls by 80\% for both ($C_{>2}$) alcohol and 3D structure systems and 42\% for oxide systems.

\section{Methods}
\label{sec:methods}

\subsection{Finetuna workflow}
\label{subsec:finetuna_wf}

\begin{figure}
\centering
\includegraphics[width=0.99\textwidth]{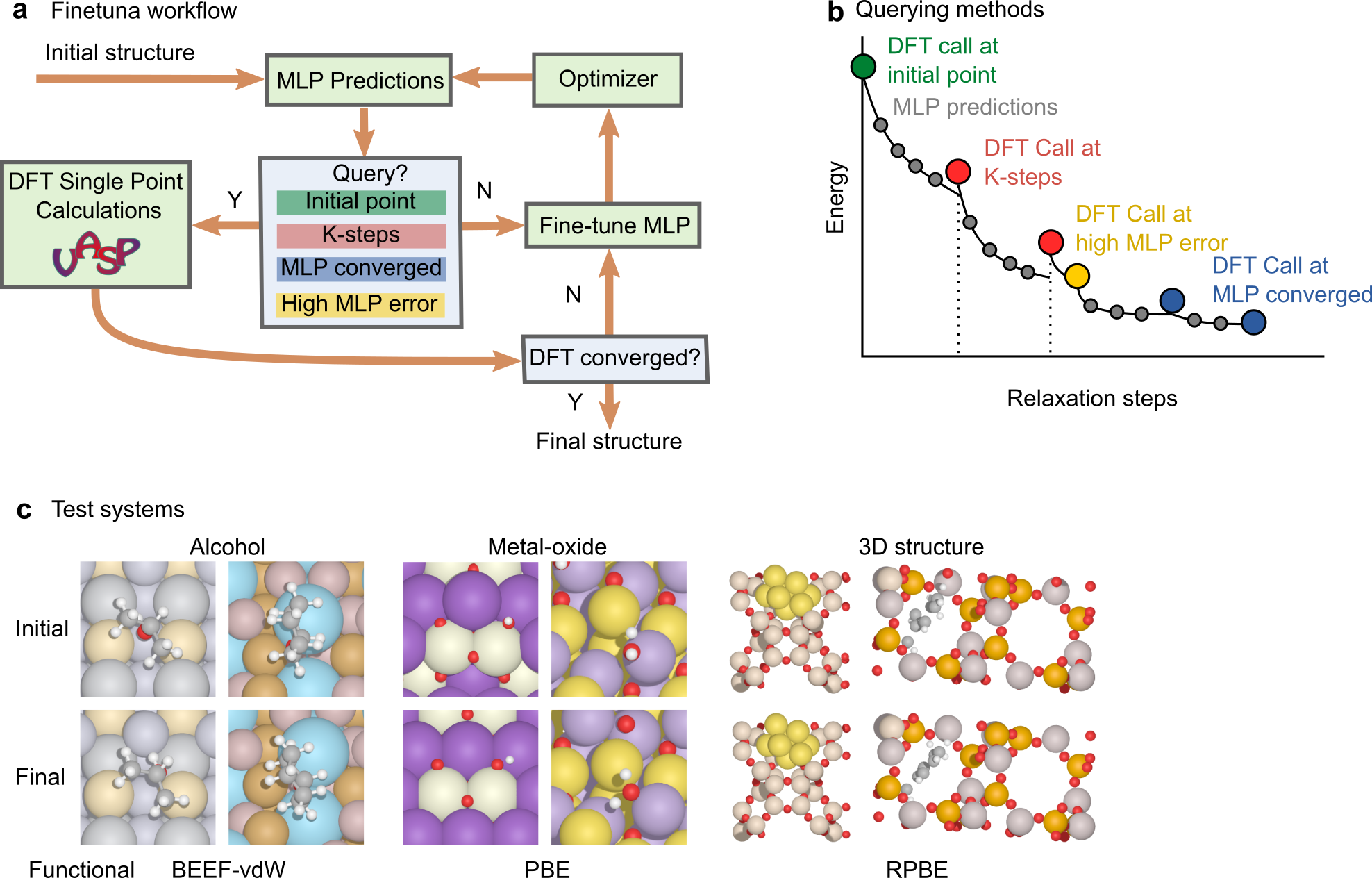}
\caption{(a) Online active learning workflow. (b) Demonstration of different querying strategies in a sample relaxation trajectory. \gls{mlp} error is based on the \gls{ml} predicted forces. (c) Sample structures of the test systems and their corresponding \gls{dft} functional used in this work.
}
\label{fig:oal_diagram}
\end{figure}

The active learning workflow accelerates geometric optimizations using a pre-trained \gls{gnn} for fast force estimations. As shown in Figure \ref{fig:oal_diagram}, the atomic structure is evaluated by the \gls{mlp}, and if a querying criterion is met, such as force convergence is reached or a step threshold is exceeded, a \gls{dft} single-point calculation will be triggered (also referred to as a ``parent call''). The \gls{dft}-calculated forces are used to fine-tune the \gls{mlp}. The optimizer uses the atomic forces to update the structure, and the \gls{mlp} is used for force prediction. Similar to other relaxation processes, the convergence criterion is based on the maximum force in the system. 

In the Finetuna workflow, many components, e.g., the \gls{ml} model, the \gls{dft} functional and software, the optimizer, and the querying strategy, can be modified to suit specific systems. The choice of the \gls{mlp} includes all the available pre-trained models, such as the GemNet model trained on \gls{oc20} dataset, the GemNet model trained on \gls{oc22} dataset, the GemNet-OC model trained on both \gls{oc20} and \gls{oc22} datasets, etc. We refer to these models as GN-OC20, GN-OC22, GNOC-OC20+22, respectively. The selection of the \gls{dft} functional depends mostly on the chemical system of interest; we discuss this further in Section \ref{subsec:chemical_systems}. We limit the \gls{dft} code to \gls{vasp} to be consistent with the \gls{oc20} and \gls{oc22} dataset \cite{Kresse1996a,Kresse1995,Kresse1994,Kresse1996b, vaspinteractive2022github}, and the optimizer to \gls{bfgs} in \gls{ase} in order to eliminate variations arising from differences between optimizers \cite{Larsen2017}. Interfaces to other \gls{dft} codes can be found in the GitHub repository \cite{almlp}, and a detailed comparison of different optimizers can be found in the prior work \cite{finetuna}. 

The querying strategy determines when a \gls{dft} calculation is performed. The four main strategies are demonstrated in Figure \ref{fig:oal_diagram}c. ``Initial point'' queries at the initial structure. ``K-steps'' determines the lower bound of querying frequency, similar to \cite{finetuna}. The `\gls{mlp} converged' strategy queries when the maximum force from \gls{ml} prediction is less than the convergence criterion. We also introduce a new strategy, labeled as ``high \gls{mlp} error'' in Figure~\ref{fig:oal_diagram}, which queries the next point whenever the previous training error fails to drop below a set threshold. The training error is defined as the L2 norm of the difference between the \gls{dft} forces and the retrained \gls{mlp} prediction. This approach is needed for high-error systems because training on a single frame may not improve the model's prediction enough to allow the framework to correctly converge.

The baseline Finetuna framework uses the GN-OC20 model as the \gls{mlp}, and queries the parent calculator at the initial point, every 30 \gls{ml} steps, or when the \gls{ml}-predicted maximum force is below the convergence criterion (in this work, 0.05 eV/\AA). 
The baseline strategy is applied to all systems discussed herein. However, the GN-OC20 model exhibits different levels of accuracy on the different types of systems tested in this work. We expect these differences to be caused by how well-represented those systems are in the \gls{oc20} training dataset. This, in turn, should affect how conservative the user ought to be in setting the Finetuna algorithm querying strategy. Systems that lie further outside the domain of the training data should require more training, and predictions by the model should be less trustworthy. Therefore, we also test different pre-trained \gls{mlp}s for the oxide systems, and apply the additional query by high \gls{mlp} error to the 3D structure systems. The detailed strategies will be further explained in Section~\ref{subsec:chemical_systems}. 


\subsection{Chemical systems}
\label{subsec:chemical_systems}

\subsubsection{Alcohol dehydrogenation}
\label{subsubsec:ipa}
Alcohol dehydrogenation has been widely explored as a means to catalytically produce and store hydrogen gas (\ce{H2}) in fuel cells. 
Modeling the dehydrogenation of relatively large ($C_{>2}$) alcohol molecules to ketones is a complex process as it involves varying possible surface intermediates at every step of the reaction network. Many useful industrial applications in catalysis involve reactions that either lead to the formation or breakdown of alcohols, such as the dehydrogenation of alcohols, the formation of fuel from \ce{CO2} reduction reactions, or the catalytic formation of plastics. Here, we assess the performance of Finetuna in modeling a two-step alcohol dehydrogenation reaction. We investigated three out-of-domain adsorbates of different sizes: $C_3$ 2-propanol or \gls{ipa} (\ce{C3H8O}), $C_4$ 2-butanol (\ce{C4H10O}), and $C_6$ cyclohexanol (\ce{C6H12O}). The surface intermediates investigated are similar to the original \gls{oc20} training dataset, but these systems are also considered out-of-domain because they are much larger than anything in the pretraining dataset. Because of this similarity, we expect training to be relatively easy and for the Finetuna algorithm to converge relatively quickly compared to more out-of-domain systems. 

For large organic molecules ($C_{>2}$), the contributions to the binding energy come from a mixture of physisorption and chemisorption. This is due to the large dipole moments inherent to these molecules which results in significant van der Waals contributions to binding energy. Modeling these molecules using \gls{dft} typically requires the addition of dispersion corrections. Here we use the \gls{beefvdw} functional to account for these dispersion interactions \cite{Wellendorff2012}. The \gls{oc20} dataset was calculated using the \gls{rpbe} functional with carbon-based adsorbates limited to $C_1$ and $C_2$ molecules without accounting for any dispersion interactions. To assess the influence of these dispersion interactions, we perform single-point \gls{dft} calculations with \gls{rpbe} and \gls{beefvdw} functionals on the initial structures of alcohol systems. We find that the force differences between the two \gls{dft} functionals are very small compared to the difference between \gls{dft} forces and \gls{mlp} forces. Details can be found in Figure~\ref{si-fig:si_rpbe_beef}.

\subsubsection{Metal-oxides}
\label{subsubsec:oxides}
Metal-oxides have been extensively studied as candidate catalysts for crucial electrochemical reactions, e.g. the \gls{oer}, \gls{orr}, and \gls{her}. However, oxide systems can be challenging to study using \gls{dft} calculations due to their complex electronic structure. The release of the \gls{oc22} dataset, which focused on metal-oxides, has provided a valuable resource to advance the understanding of metal-oxide systems. Nevertheless, it is important to note that the \gls{oc22} dataset represents only a small subset of the extensive design space of oxides, underscoring the need for efficient exploration of these systems. Metal-oxide systems tested in this work are taken from the \gls{oc22} validation set. These are also adsorbate/slab configurations similar in form to training data in the \gls{oc20} dataset. These systems often exhibit spin ordering, so spin-polarized calculations are generally required. In addition, a Hubbard U correction is applied to certain elements in these systems to improve the description of localized electron states\cite{Jain2013}. We suspect this can result in a potential energy surface with multiple nearby local minima, and thus give different outcomes when performing geometric optimizations \cite{Meredig1951}. For the selected testing systems, we used the same \gls{dft} settings from the \gls{oc22} dataset \cite{oc22}. 

Evaluating pre-trained graph models like GemNet on oxide systems results in higher errors than non-oxide materials, due to the complexity of oxides. 
Training on both oxide and non-oxide materials improves the model accuracy, but energy and force prediction on oxide materials is still a more difficult task. 
Graph model featurization schemes may fail to capture some complex properties that are more common in, or specific to, oxide materials, such as magnetic effects and long-range charge effects \cite{Akpabio2021}.
Unlike metallic systems, semiconductors can have essentially identical structures, but relax to significantly different final geometries due to long-range effects of the number of shared electrons across the entire system, or exhibiting different magnetic configurations of the same structure.
It has been shown that the GemNet-OC model trained only on \gls{oc20} dataset performs especially poorly on \gls{oc22} prediction tasks, e.g. the force MAE is 0.384 eV/{\AA} \cite{oc22}. 
Even when given many training examples like those in the \gls{oc22} dataset, the accuracy of \gls{gnn}s like GemNet on oxide systems does not approach the accuracy on non-oxide systems. 
As shown in the \gls{ocp} leaderboard, the best GemNet model for out-of-distribution force predictions results in a force \gls{mae} of 0.031 eV/{\AA} for \gls{oc22}, versus 0.023 eV/{\AA} force MAE for \gls{oc20} at the time of writing \cite{oc20, oc22}. 
In addition to the baseline Finetuna strategy that uses GN-OC20, we also test the performance of GN-OC22 and GNOC-OC20+22 as the underlying \gls{mlp}, due to the anticipated difficulty of this task.
The interaction between magnetic spins in systems with spin polarization is one of the contributing factors to the complexity of oxide materials. 
Spin polarization effects are typically long-range. 
However, \gls{gnn} models like GemNet are developed under the assumption that local interactions dominate \cite{oc22}.

\subsubsection{3D structures}
\label{subsubsec:zeolites}
Zeolites and \gls{mof}s are two common materials with porous structures. They have gained significant attention in recent decades due to their versatile properties and wide-ranging applications in gas separation, water treatment, and catalysis \cite{VanSpeybroeck2015, Lee2009, Yang2019}. One advantage of these 3D structures as catalysts is their tunability. By modifying the composition, structure, and pore size, their selectivity and performance can be calibrated for specific applications \cite{Bacariza2019, Pascanu2019}. 

The test 3D structures cover a wide range of systems, including copper-doped zeolites, \gls{mof}, and zeolites with a variety of adsorbates such as gold nanoparticles and aromatics. These catalysts are modeled very differently from \gls{oc20}-like metal surfaces. Rather than adsorbing  onto the surface of a dimensionally confined slab structure, which is the core focus of the \gls{oc20} dataset, these structures are fully three-dimensional, and adsorbates, if present, are incorporated into pores in the 3D structure. Structural relaxations then take place throughout the simulation cell (rather than only in a few atomic layers near the surface/adsorbate). Examples of such complex structures are not found within the \gls{oc20} training dataset. While there may be numerous examples of certain environments containing similar metal-oxygen configurations, there should be many differences due to the addition of three-dimensional surroundings for most atomic neighborhoods. The metal shells of these 3D structures are also effectively made up of metal-oxide configurations, adding to the difficulty. We anticipate this kind of structure to be very difficult to adapt to for a \gls{gnn} model like GemNet, trained only on \gls{oc20}-like structures.

To compensate for this, we test another, more conservative, querying strategy for Finetuna, in addition to the baseline approach. We measure the mean error of the retrained forces after each parent call during the Finetuna loop; by taking the L2 norm of the difference between the parent forces and the retrained model prediction. If the retrained mean force error (after fine-tuning) is still above some threshold, this signals that training was insufficient, and the parent \gls{dft} call is triggered on the next step of the relaxation as well. This process repeats at every parent call until the model error drops below a threshold (in this case 0.05 eV/{\AA}). This approach aims to ensure that the model is sufficiently trained on the new system to make force predictions that will take the relaxation in a reasonable direction to reduce the forces and energy.

\section{Results \& Discussion}
\label{sec:result}



We selected 27 alcohol-to-ketone systems, 10 oxide systems, and 81 3D structure systems and performed both \gls{vasp} \gls{bfgs} and Finetuna relaxations. The systems that experienced convergence issues with \gls{vasp} \gls{bfgs} and the systems that failed due to memory limitations are excluded from this section. A full list of the tested systems and results can be found in the Supplementary Information. The size of the chemical systems in the experiment varied from 50 to 250 atoms. We report the energy difference between the \gls{vasp} \gls{bfgs} and Finetuna relaxations in eV/atom to facilitate comparison.

\begin{figure}[ht]
\centering
\includegraphics[width=0.99\textwidth]{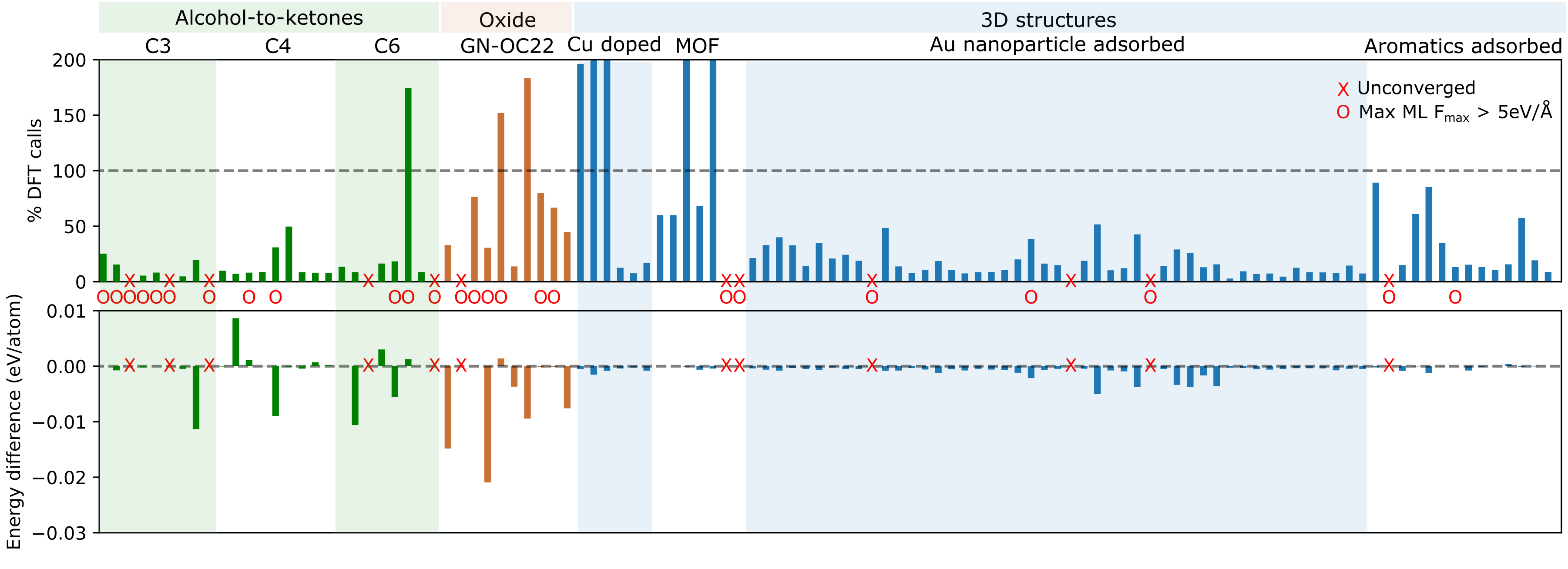}
\caption{Finetuna performance compared to VASP BFGS on (a) alcohols, (b) oxides, and (c) zeolite systems. The red `X' represents the unconverged Finetuna relaxations, and the red `O' labels the systems where a high \gls{mlp} maximum force is observed along the relaxation trajectory. Comprehensive information regarding each individual system can be found in the Supplementary Information (Table~\ref{si-tab:ipa_summary}, Table~\ref{si-tab:oxide_summary}, and Table~\ref{si-tab:zeolite_summary}). }
\label{fig:bar_summary}
\end{figure}

A summary of the overall performance of Finetuna compared to \gls{vasp} \gls{bfgs} can be found in Table~\ref{tab:summary_table}, and a plot of the individual runs is shown in Figure~\ref{fig:bar_summary}. For each class of systems, we evaluate the efficiency of the Finetuna workflow by calculating the ratio between the total number of \gls{dft} calls with Finetuna relaxation and that with \gls{vasp} \gls{bfgs}. This metric is referred to as the percentage \gls{dft} calls and represents the overall saving of \gls{dft} calculations. The unconverged Finetuna runs are labeled in red crosses in Figure~\ref{fig:bar_summary}. We found that none of the metal-oxide systems converged with the baseline GN-OC20 model. This is not surprising due to the lack of oxide systems and spin polarization information in the \gls{oc20} training data. For simplicity, the results from Finetuna with GN-OC22 model are shown in Figure~\ref{fig:bar_summary}, and the comparisons with other models can be found in the supporting information Figure~\ref{si-fig:si_oxide}. In general, Finetuna underperforms on both accuracy and speed when compared to the results in Ref. \citenum{finetuna} across all three system types. We believe this is the result of a significant domain shift, which should make realigning the \gls{gnn} model with fine-tuning slower and less accurate. We consider magnetic effects and 3D structural effects to be the most significant causes of domain shift. By increasing the conservativeness in the Finetuna algorithm through repeated training when errors are high, the overall parent calls can be brought below 20\% for alcohol-to-ketones and the 3D structures, and 58\% for oxide systems.
\begin{table}[ht]
\caption{Summary of the Finetuna performance on alcohol, oxide, and 3D structure systems. The baseline querying strategy calls a \gls{dft} calculation at the initial point, every 30 \gls{mlp} steps, and when \gls{mlp} predicted forces meet the convergence criterion.}
\label{tab:summary_table}
\resizebox{\columnwidth}{!}{

\begin{tabular}{|l|l|l|l|l|l|}
\hline
System                  & ML model & Querying strategy                                                   & Functional           & \% Converged & \% DFT calls \\ \hline
Alcohols & GN-OC20       & Baseline & BEEF-vdW & 81\% & 18\% \\ \hline
\multirow{3}{*}{Oxides} & GN-OC20    & \multirow{3}{*}{Baseline}                                           & \multirow{3}{*}{PBE} & 0\%          & N/A          \\ \cline{2-2} \cline{5-6} 
    & GN-OC22       &          &          & 90\% & 58\% \\ \cline{2-2} \cline{5-6} 
    & GNOC-OC20+22 &          &          & 70\% & 36\% \\ \hline
Zeolite                 & GN-OC20  & \begin{tabular}[c]{@{}l@{}}Baseline\\ + High MLP error\end{tabular} & RPBE                 & 92\%         & 19\%         \\ \hline
\end{tabular}
}
\end{table}

\begin{figure}[ht]
\centering
\includegraphics[width=0.99\textwidth]{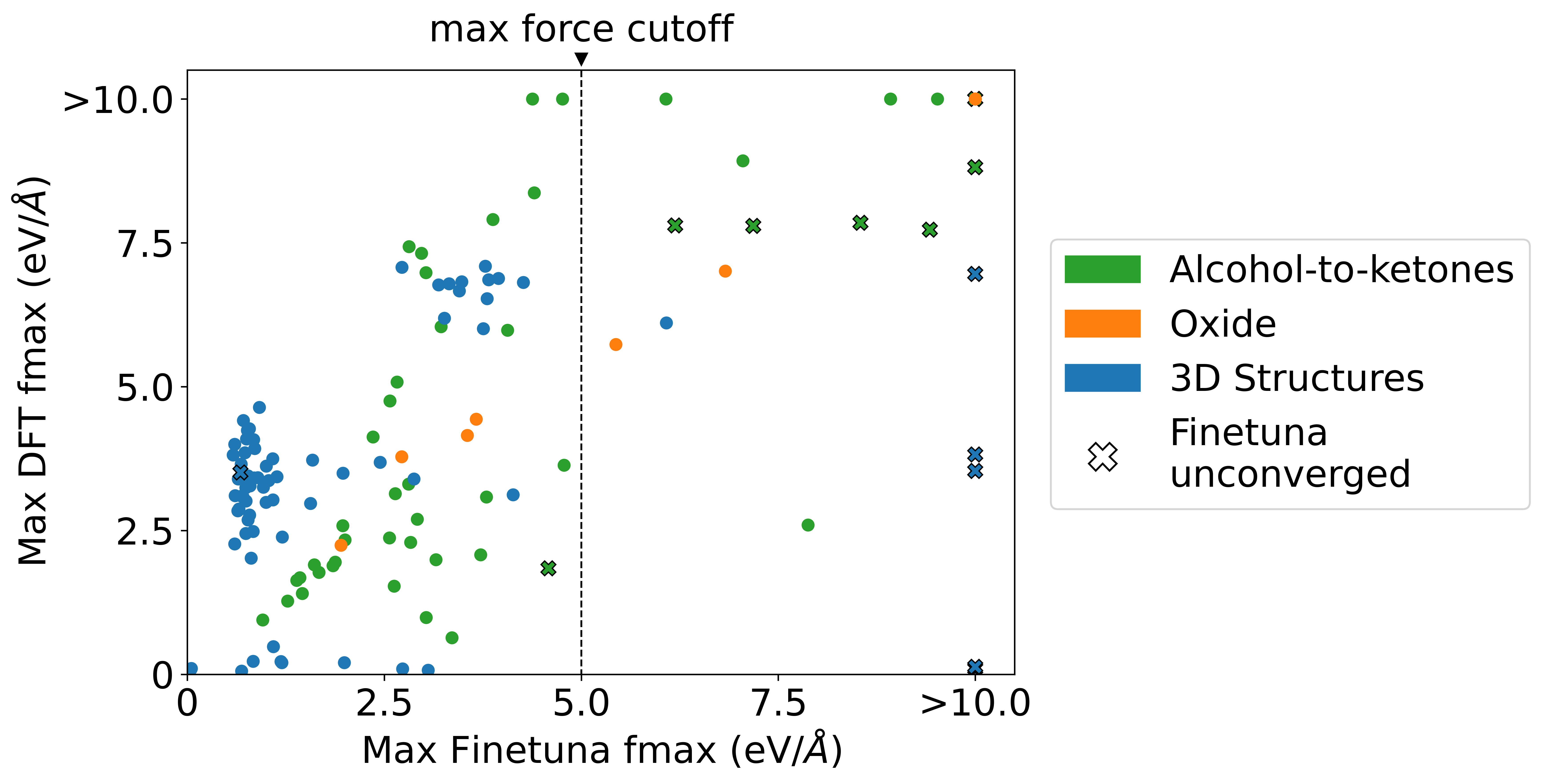}
\caption{Performance metrics plotted against the maximum force at a parent call along each Finetuna trajectory. Each point corresponds to a separate relaxation trajectory. Along the $x$-axis is the maximum force error for a parent call, measured by taking the L2 norm of the difference between each force vector, and taking the mean of those over the whole system at that point.}
\label{fig:error_analysis_plot}
\end{figure}

Across alcohol-to-ketones, metal-oxides, and 3D structure systems, there are a number of cases where Finetuna does not converge. As shown in Figure~\ref{fig:error_analysis_plot}, most of the unconverged cases have observed high \gls{ml} predicted forces (> 5.0 eV/{\AA}) across the entire relaxation trajectory. This implies that for trajectories where the \gls{mlp} made a large force prediction, they generally should not be trusted to converge, and should be terminated early. In addition, we would expect that (provided initial structure guesses are reasonable), the occurrence of such a large force would indicate that the system has moved to a less realistic region of the configuration space and any local minimum it finds would likely be less physically meaningful.
The only exceptions to this are with a few of the alcohol-to-ketones systems, where the trajectory slowly converged (exceeding 100\% of the \gls{dft} parent calls) despite the high error.
Interestingly, these systems do not appear to find significantly different local minima from the original \gls{dft} relaxation, so in these cases, it would appear that Finetuna took a somewhat roundabout route to find a very similar result.

\begin{figure}[ht]
\centering
\includegraphics[width=0.99\textwidth]{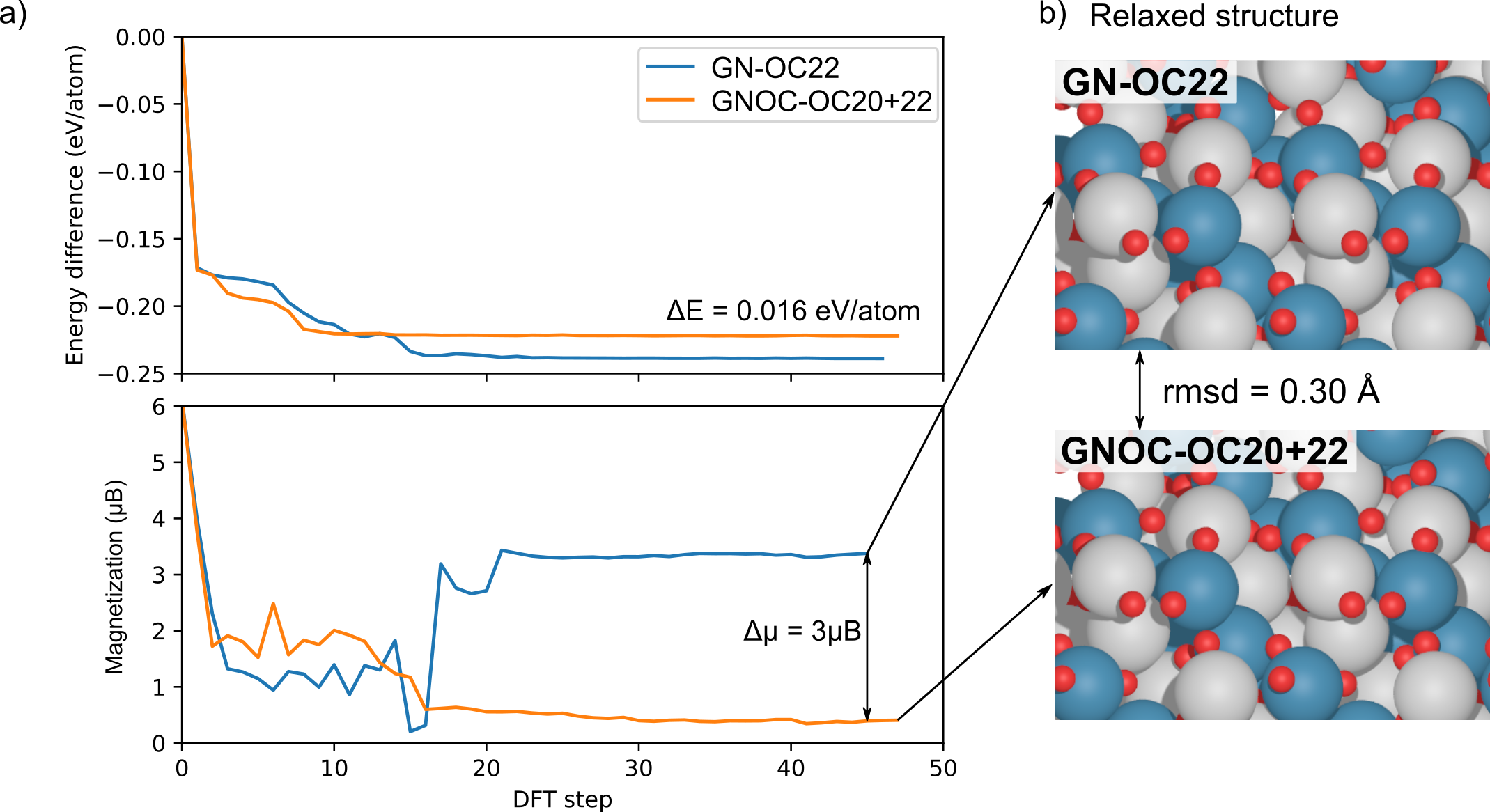}
\caption{Detailed analysis of the \ce{AgIrO4} system. (a) Change of \gls{dft} energy and magnetization along the relaxation. (b) Comparison of the relaxed structure from two Finetuna relaxations. }
\label{fig:spin_analysis}
\end{figure}

For the oxide systems, we found that the potential energy surfaces are generally more complicated, as evidenced by the discrepancies in the relaxed energy from the two \gls{dft} calculations (\gls{vasp} \gls{bfgs} and \gls{vasp} \gls{cg} in Figure~\ref{si-fig:si_oxide}). The total magnetization ($M$) is given by the net spin of the electrons in the system:
\begin{equation}
M=\mu_{\text{B}}\int\left ( n_\uparrow -n_\downarrow \right )d^3r,
\end{equation}
where $\mu_{\text{B}}$ is the Bohr magneton, and $n_\uparrow$ and $n_\downarrow$ represent spin densities. For our purposes, $M$ represents the extent of spin polarization in a given system. Figure~\ref{fig:spin_analysis} shows a case where relaxation with two different models led to nearly identical sets of atomic positions but considerably different magnetizations and energies.
The spin polarization adds another degree of freedom to the local minimization process and breaks the one-to-one mapping between atomic structure and potential energy presumed by GemNet. Additionally, as previously mentioned, the \gls{ml} model we use in this work sets a local cutoff when representing the atomic structure, and is unable to capture long-range interaction from magnetic effects. This suggests that the \gls{ml} potential and the performance of the algorithm may be compromised when applied to spin-polarized systems, and highlights the importance of developing \gls{mlp}s, such as CHGNET \cite{Deng2023}, that effectively capture magnetic information for these applications.

\section{Conclusion}
\label{sef:conclusion}
In this study, we conducted experiments to assess the performance of the Finetuna workflow on various chemical systems including alcohol-to-ketones, metal-oxides, and \gls{mof}s. Our findings revealed that, while showing promising results in the original domain (\gls{oc20}-like systems) with a significant reduction of \gls{dft} calls by 90\%, Finetuna was less performant on these out-of-domain systems. The results show that we were able to reduce the number of DFT calls by 85\% for alcohol-to-ketones systems, 49\% for oxide systems, and 82\% for \gls{mof} systems, with a few unconverged cases. We believe that the complex electronic interactions in these chemical systems, particularly the oxides with spin polarization, limit Finetuna's performance. The long-range magnetic interaction introduces more complexity to the potential energy surface. While this has affected the efficiency of the Finetuna workflow, it also presents opportunities for further exploration. We have established a heuristic criterion to predict the success of a Finetuna run. Specifically, if the maximum force predicted by the \gls{ml} model exceeds 5 eV/\AA, we recommend using a different initialization strategy, or terminating the simulation and performing \gls{dft} calculations instead.

As the accuracy of \gls{ml} models improves and more material datasets become available, we believe that the Finetuna workflow can be applied to a broader range of chemical systems. We anticipate the rapid emergence of new \gls{ml} models and look forward to their integration into the workflow. However, to effectively utilize these new models, further research is needed to explore transfer learning strategies and develop fine-tuning techniques specific to the models. The choice of a pre-trained model with suitable architectural features and trained on relevant data can significantly impact the model's performance and generalizability. Investigating out-of-domain detection methods can enhance the robustness of our approach. Additionally, the inclusion of uncertainty quantification in \gls{ml} models would be valuable, as it could be leveraged as a querying strategy to guide the selection of informative data points for further exploration. These future directions hold promise for advancing the efficiency and accuracy of material discovery applications. The active learning and finetuning approach has the potential to greatly accelerate molecular simulations, leading to the development of more efficient and sustainable materials with a wide range of applications.

\section{Acknowledgements}
\label{sef:acknowledgements}
The authors acknowledge support from Meta Platforms, Inc. via the Open Catalyst Project collaboration.

\section{Data Availability Statement}
\label{sec:githubrepo}

All data that support the findings of this study are included within the Github repository at {\url{https://github.com/ulissigroup/finetuna-2023-manuscript}}.
    
\bibliography{reference}
\end{document}


\maketitle

\tableofcontents

\clearpage
\label{sec:si}

\FloatBarrier
\section{Oxide results}
\label{subsec:oxide_result}
\begin{figure}[ht]
\centering
\includegraphics[width=0.99\textwidth]{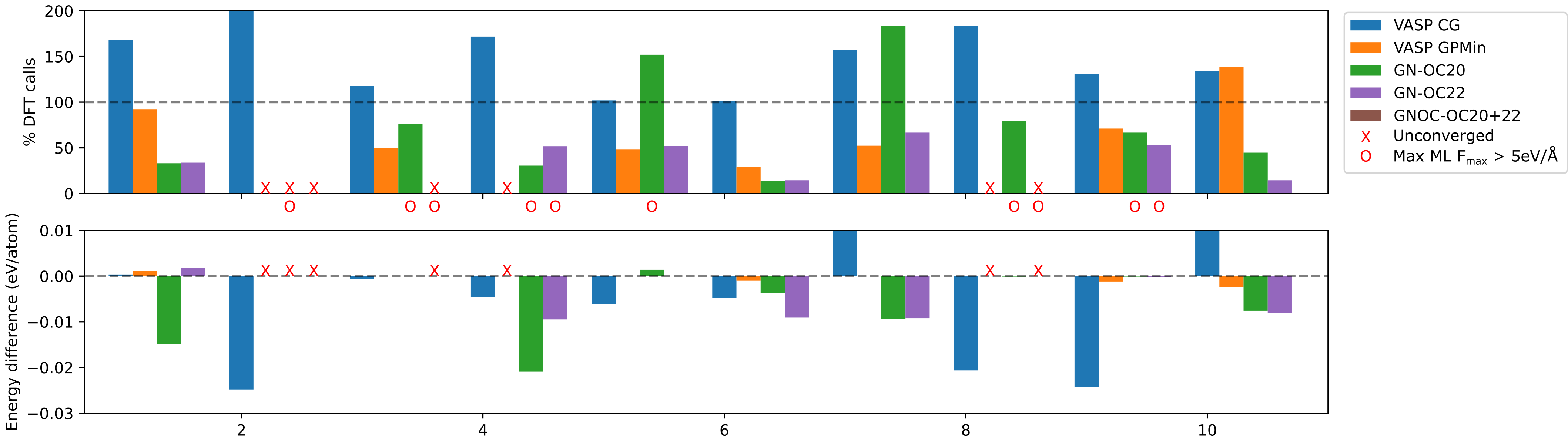}
\caption{DFT and Finetuna performance compared to VASP BFGS on metal oxide systems. GN-OC20 experiments are not included in the figure because none of them converged.}
\label{fig:si_oxide}
\end{figure}

The outcomes of the experimentation involving six distinct calculation methods applied to the metal oxide systems are summarized in Figure~\ref{fig:si_oxide}. The \gls{vasp} \gls{bfgs} method serves as the baseline, \gls{vasp} \gls{cg} uses the built-in \gls{cg} optimization method in \gls{vasp}, \gls{vasp} \gls{gpmin} uses Gaussian processes to model the potential energy surface and is an optimization method that is built in to \gls{ase}, and the rest are the Finetuna experiments with GN-OC20, GN-OC22, and GNOC-OC20+22 respectively. None of the Finetuna runs with GN-OC20 as the model converged, and is therefore not included in the plot.

\begin{table}
\caption{Summary of DFT and Finetuna performance on oxide systems. The baseline querying strategy calls a \gls{dft} calculation at the initial point, every 30 \gls{mlp} steps, and when \gls{mlp} predicted forces meet the convergence criterion.}
\label{tab:si-oxide}
\begin{tabular}{|c|c|c|c|}
\hline
System & Algorithm & \% Converged & \% DFT call \\ \hline
\multirow{6}{*}{Oxide} & VASP BFGS & 100\% & 100\% \\ \cline{2-4} 
 & VASP CG & 100\% & 154\% \\ \cline{2-4} 
 & VASP GPMin & 70\% & 70\% \\ \cline{2-4} 
 & GN-OC20 & 0\% & N/A \\ \cline{2-4} 
 & GN-OC22 & 90\% & 58\% \\ \cline{2-4} 
 & GNOC-OC20+22 & 70 & 36\% \\ \hline
\end{tabular}
\end{table}

\FloatBarrier
\section{Functional comparison}
\label{subsec:result}
\begin{figure}[ht]
\centering
\includegraphics[width=0.99\textwidth]{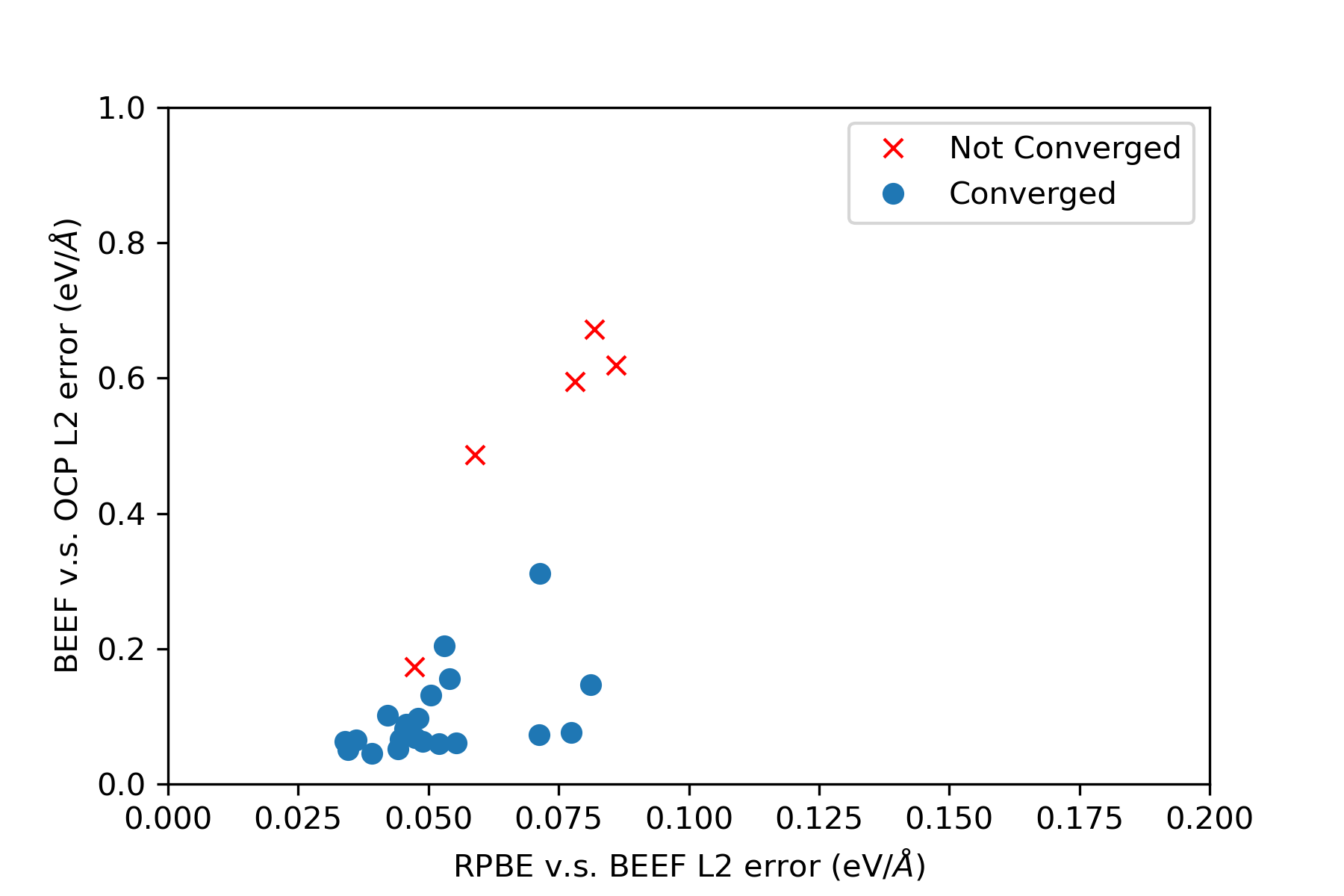}
\caption{Comparison between RPBE, BEEF-vdW, and pre-trained OCP on alcohol systems.}
\label{fig:si_rpbe_beef}
\end{figure}

The initial structure of the selected alcohol systems is calculated with \gls{rpbe}, \gls{beefvdw} functionals, and the GN-OC20 model. For each system, the L2 norm of the differences between the two \gls{dft} forces, and the \gls{dft} and \gls{mlp} obtained from these methods is plotted in Figure~\ref{fig:si_rpbe_beef}. The systems with unconverged Finetuna runs are labeled as `x's. As shown in the figure, the differences between \gls{dft} forces of all systems are less than 0.1 eV/{\AA}, whereas the differences between \gls{dft} with \gls{beefvdw} functional and the GN-OC20 model vary. Particularly, when the discrepancy between \gls{dft} and \gls{mlp} is high, i.e.: above 0.4 eV/{\AA}, the Finetuna run tends to fail. 

\FloatBarrier
\section{Similarity Mapping}
\label{subsec:similarity}

\begin{figure}[ht]
\centering
\includegraphics[width=0.99\textwidth]{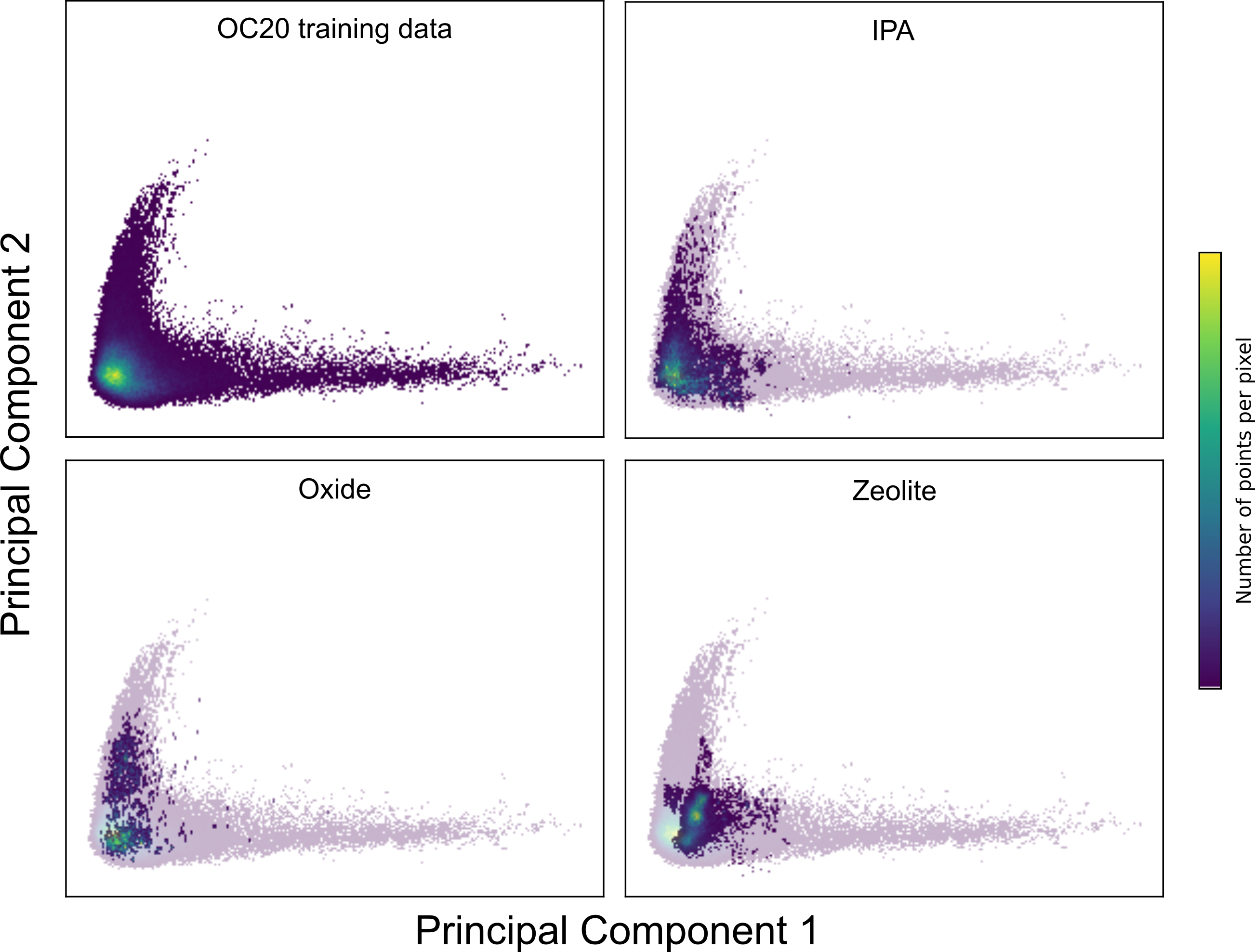}
\caption{\gls{pca} maps of each of the system types.}
\label{fig:si_pca}
\end{figure}

To better understand where our test systems sit relative to the distribution of \gls{oc20} training data, we perform latent space analysis with \gls{pca}, a dimensionality reduction method. The atomic embeddings are extracted from the latent space of the final interaction block of the GemNet model. These descriptors provide a 256-dimensional vector on a per-atom basis, which should describe the atomic neighborhood of each atom in each frame of the training set. 
We plot the atomic environment on a per-atom basis for these maps because a high predictive error on a single atom should be sufficient to disrupt an entire relaxation. The density of each map gives us a qualitative understanding that the alcohol-to-ketone systems should be most in-domain, while the zeolite systems appear to be furthest out of the domain. However, this type of plot does not give us a quantitative measure with which to set a threshold.

\FloatBarrier
\section{Experiment results}
\label{subsec:overfitting}
\begin{table}[ht]
\caption{VASP and Finetuna result for alcohol systems.}
\label{tab:ipa_summary}
\resizebox{\columnwidth}{!}{
\begin{tabular}{|r|r|r|r|r|r|}
\hline
\multicolumn{1}{|c|}{System name} &
  \multicolumn{1}{c|}{Algorithm} &
  \multicolumn{1}{c|}{\begin{tabular}[c]{@{}c@{}}\% DFT\\calls\end{tabular}} &
  \multicolumn{1}{c|}{\begin{tabular}[c]{@{}c@{}}No. DFT\\calls\end{tabular}} &
  \multicolumn{1}{c|}{Converged} &
  \multicolumn{1}{c|}{\begin{tabular}[c]{@{}c@{}}Relaxed energy\\ difference (eV)\end{tabular}} \\ \hline
\multirow{2}{*}{mp-1029784\_33-2\_0.0405\_H6C3O\_2\_adslab} & VASP BFGS & 100.00 & 198 & True & 0.0000 \\ \cline{2-6} 
 & GN-OC20 & 25.25 & 50 & True & 0.0000 \\ \hline
\multirow{2}{*}{mp-11300\_320\_0.0500\_H6C3O\_31\_adslab} & VASP BFGS & 100.00 & 71 & True & 0.0000 \\ \cline{2-6} 
 & GN-OC20 & 15.49 & 11 & True & -0.0007 \\ \hline
\multirow{2}{*}{mp-989621\_023\_0.0311\_H6C3O\_14\_adslab} & VASP BFGS & 100.00 & 149 & True & 0.0000 \\ \cline{2-6} 
 & GN-OC20 & 0.00 & N/A & False & N/A \\ \hline
\multirow{2}{*}{mp-1029784\_33-2\_0.0405\_H7C3O\_2\_adslab} & VASP BFGS & 100.00 & 253 & True & 0.0000 \\ \cline{2-6} 
 & GN-OC20 & 5.53 & 14 & True & -0.0002 \\ \hline
\multirow{2}{*}{mp-11300\_320\_0.0500\_H7C3O\_0\_adslab} & VASP BFGS & 100.00 & 97 & True & 0.0000 \\ \cline{2-6} 
 & GN-OC20 & 8.25 & 8 & True & 0.0000 \\ \hline
\multirow{2}{*}{mp-989621\_023\_0.0311\_H7C3O\_38\_adslab} & VASP BFGS & 100.00 & 105 & True & 0.0000 \\ \cline{2-6} 
 & GN-OC20 & 0.00 & N/A & False & N/A \\ \hline
\multirow{2}{*}{mp-1029784\_33-2\_0.0405\_H8C3O\_31\_adslab} & VASP BFGS & 100.00 & 250 & True & 0.0000 \\ \cline{2-6} 
 & GN-OC20 & 4.80 & 12 & True & -0.0005 \\ \hline
\multirow{2}{*}{mp-11300\_320\_0.0500\_H8C3O\_23\_adslab} & VASP BFGS & 100.00 & 82 & True & 0.0000 \\ \cline{2-6} 
 & GN-OC20 & 19.51 & 16 & True & -0.0113 \\ \hline
\multirow{2}{*}{mp-989621\_023\_0.0311\_H8C3O\_2\_adslab} & VASP BFGS & 100.00 & 115 & True & 0.0000 \\ \cline{2-6} 
 & GN-OC20 & 0.00 & N/A & False & N/A \\ \hline
\multirow{2}{*}{mp-21035\_210\_0.0833\_H10C4O\_16\_adslab} & VASP BFGS & 100.00 & 162 & True & 0.0000 \\ \cline{2-6} 
 & GN-OC20 & 9.88 & 16 & True & -0.0001 \\ \hline
\multirow{2}{*}{mp-30451\_320\_0.0333\_H10C4O\_52\_adslab} & VASP BFGS & 100.00 & 350 & True & 0.0000 \\ \cline{2-6} 
 & GN-OC20 & 7.14 & 25 & True & 0.0087 \\ \hline
\multirow{2}{*}{mp-644278\_100\_0.1667\_H10C4O\_3\_adslab} & VASP BFGS & 100.00 & 451 & True & 0.0000 \\ \cline{2-6} 
 & GN-OC20 & 8.20 & 37 & True & 0.0012 \\ \hline
\multirow{2}{*}{mp-21035\_210\_0.0833\_H8C4O\_12\_adslab} & VASP BFGS & 100.00 & 91 & True & 0.0000 \\ \cline{2-6} 
 & GN-OC20 & 8.79 & 8 & True & -0.0000 \\ \hline
\multirow{2}{*}{mp-30451\_320\_0.0333\_H8C4O\_12\_adslab} & VASP BFGS & 100.00 & 152 & True & 0.0000 \\ \cline{2-6} 
 & GN-OC20 & 30.92 & 47 & True & -0.0090 \\ \hline
\multirow{2}{*}{mp-644278\_100\_0.1667\_H8C4O\_6\_adslab} & VASP BFGS & 100.00 & 155 & True & 0.0000 \\ \cline{2-6} 
 & GN-OC20 & 49.68 & 77 & True & -0.0001 \\ \hline
\multirow{2}{*}{mp-21035\_210\_0.0833\_H9C4O\_10\_adslab} & VASP BFGS & 100.00 & 106 & True & 0.0000 \\ \cline{2-6} 
 & GN-OC20 & 8.49 & 9 & True & -0.0004 \\ \hline
\multirow{2}{*}{mp-30451\_320\_0.0333\_H9C4O\_40\_adslab} & VASP BFGS & 100.00 & 111 & True & 0.0000 \\ \cline{2-6} 
 & GN-OC20 & 8.11 & 9 & True & 0.0007 \\ \hline
\multirow{2}{*}{mp-644278\_100\_0.1667\_H9C4O\_13\_adslab} & VASP BFGS & 100.00 & 182 & True & 0.0000 \\ \cline{2-6} 
 & GN-OC20 & 7.69 & 14 & True & 0.0002 \\ \hline
\multirow{2}{*}{mp-1205604\_331\_0.0208\_H10C6O\_38\_adslab} & VASP BFGS & 100.00 & 243 & True & 0.0000 \\ \cline{2-6} 
 & GN-OC20 & 13.58 & 33 & True & -0.0000 \\ \hline
\multirow{2}{*}{mp-1215471\_110\_0.2500\_H10C6O\_22\_adslab} & VASP BFGS & 100.00 & 199 & True & 0.0000 \\ \cline{2-6} 
 & GN-OC20 & 8.54 & 17 & True & -0.0106 \\ \hline
\multirow{2}{*}{mp-1223339\_210\_0.0833\_H10C6O\_32\_adslab} & VASP BFGS & 100.00 & 114 & True & 0.0000 \\ \cline{2-6} 
 & GN-OC20 & 0.00 & N/A & False & N/A \\ \hline
\multirow{2}{*}{mp-1205604\_331\_0.0208\_H11C6O\_48\_adslab} & VASP BFGS & 100.00 & 238 & True & 0.0000 \\ \cline{2-6} 
 & GN-OC20 & 16.39 & 39 & True & 0.0030 \\ \hline
\multirow{2}{*}{mp-1215471\_110\_0.2500\_H11C6O\_14\_adslab} & VASP BFGS & 100.00 & 131 & True & 0.0000 \\ \cline{2-6} 
 & GN-OC20 & 18.32 & 24 & True & -0.0056 \\ \hline
\multirow{2}{*}{mp-1223339\_210\_0.0833\_H11C6O\_18\_adslab} & VASP BFGS & 100.00 & 110 & True & 0.0000 \\ \cline{2-6} 
 & GN-OC20 & 174.55 & 192 & True & 0.0012 \\ \hline
\multirow{2}{*}{mp-1215471\_110\_0.2500\_H12C6O\_19\_adslab} & VASP BFGS & 100.00 & 104 & True & 0.0000 \\ \cline{2-6} 
 & GN-OC20 & 8.65 & 9 & True & -0.0000 \\ \hline
\multirow{2}{*}{mp-1223339\_210\_0.0833\_H12C6O\_40\_adslab} & VASP BFGS & 100.00 & 120 & True & 0.0000 \\ \cline{2-6} 
 & GN-OC20 & 0.00 & N/A & False & N/A \\ \hline
\end{tabular}
}
\end{table}

\begin{table}[ht]
\vspace{-5\baselineskip}
\caption{VASP and Finetuna with different optimizers and different pre-trained model results for metal oxides.}
\label{tab:oxide_summary}
\resizebox{0.9\columnwidth}{!}{
\begin{tabular}{|r|r|r|r|r|r|}
\hline
\multicolumn{1}{|c|}{System name} &
  \multicolumn{1}{c|}{Algorithm} &
  \multicolumn{1}{c|}{\begin{tabular}[c]{@{}c@{}}\% DFT\\calls\end{tabular}} &
  \multicolumn{1}{c|}{\begin{tabular}[c]{@{}c@{}}No. DFT\\calls\end{tabular}} &
  \multicolumn{1}{c|}{Converged} &
  \multicolumn{1}{c|}{\begin{tabular}[c]{@{}c@{}}Relaxed energy\\ difference (eV)\end{tabular}} \\ \hline
\multirow{6}{*}{AgIrO4} & VASP BFGS & 100.00 & 142 & True & 0.0000 \\ \cline{2-6} 
 & VASP CG & 168.31 & 239 & True & 0.0004 \\ \cline{2-6} 
 & VASP GPMin & 92.25 & 131 & True & 0.0011 \\ \cline{2-6} 
 & GN-OC20 & 0.00 & N/A & False & N/A \\ \cline{2-6} 
 & GN-OC22 & 33.10 & 47 & True & -0.0148 \\ \cline{2-6} 
 & GNOC-OC20+22 & 33.80 & 48 & True & 0.0019 \\ \hline
\multirow{6}{*}{IrAuO4} & VASP BFGS & 100.00 & 136 & True & 0.0000 \\ \cline{2-6} 
 & VASP CG & 217.65 & 296 & True & -0.0248 \\ \cline{2-6} 
 & VASP GPMin & 0.00 & N/A & False & N/A \\ \cline{2-6} 
 & GN-OC20 & 0.00 & N/A & False & N/A \\ \cline{2-6} 
 & GN-OC22 & 0.00 & N/A & False & N/A \\ \cline{2-6} 
 & GNOC-OC20+22 & 0.00 & N/A & False & N/A \\ \hline
\multirow{6}{*}{IrRhO4} & VASP BFGS & 100.00 & 34 & True & 0.0000 \\ \cline{2-6} 
 & VASP CG & 117.65 & 40 & True & -0.0007 \\ \cline{2-6} 
 & VASP GPMin & 50.00 & 17 & True & -0.0000 \\ \cline{2-6} 
 & GN-OC20 & 0.00 & N/A & False & N/A \\ \cline{2-6} 
 & GN-OC22 & 76.47 & 26 & True & -0.0000 \\ \cline{2-6} 
 & GNOC-OC20+22 & 0.00 & N/A & False & N/A \\ \hline
\multirow{6}{*}{Li2FeO3} & VASP BFGS & 100.00 & 85 & True & 0.0000 \\ \cline{2-6} 
 & VASP CG & 171.76 & 146 & True & -0.0046 \\ \cline{2-6} 
 & VASP GPMin & 0.00 & N/A & False & N/A \\ \cline{2-6} 
 & GN-OC20 & 0.00 & N/A & False & N/A \\ \cline{2-6} 
 & GN-OC22 & 30.59 & 26 & True & -0.0209 \\ \cline{2-6} 
 & GNOC-OC20+22 & 51.76 & 44 & True & -0.0095 \\ \hline
\multirow{6}{*}{Li4SeO5} & VASP BFGS & 100.00 & 52 & True & 0.0000 \\ \cline{2-6} 
 & VASP CG & 101.92 & 53 & True & -0.0061 \\ \cline{2-6} 
 & VASP GPMin & 48.08 & 25 & True & 0.0001 \\ \cline{2-6} 
 & GN-OC20 & 0.00 & N/A & False & N/A \\ \cline{2-6} 
 & GN-OC22 & 151.92 & 79 & True & 0.0014 \\ \cline{2-6} 
 & GNOC-OC20+22 & 51.92 & 27 & True & -0.0000 \\ \hline
\multirow{6}{*}{MnAuO2} & VASP BFGS & 100.00 & 145 & True & 0.0000 \\ \cline{2-6} 
 & VASP CG & 101.38 & 147 & True & -0.0048 \\ \cline{2-6} 
 & VASP GPMin & 28.97 & 42 & True & -0.0010 \\ \cline{2-6} 
 & GN-OC20 & 0.00 & N/A & False & N/A \\ \cline{2-6} 
 & GN-OC22 & 13.79 & 20 & True & -0.0037 \\ \cline{2-6} 
 & GNOC-OC20+22 & 14.48 & 21 & True & -0.0091 \\ \hline
\multirow{6}{*}{NaScO2} & VASP BFGS & 100.00 & 42 & True & 0.0000 \\ \cline{2-6} 
 & VASP CG & 157.14 & 66 & True & 0.0170 \\ \cline{2-6} 
 & VASP GPMin & 52.38 & 22 & True & 0.0000 \\ \cline{2-6} 
 & GN-OC20 & 0.00 & N/A & False & N/A \\ \cline{2-6} 
 & GN-OC22 & 183.33 & 77 & True & -0.0094 \\ \cline{2-6} 
 & GNOC-OC20+22 & 66.67 & 28 & True & -0.0092 \\ \hline
\multirow{6}{*}{Rb2CeO3} & VASP BFGS & 100.00 & 84 & True & 0.0000 \\ \cline{2-6} 
 & VASP CG & 183.33 & 154 & True & -0.0207 \\ \cline{2-6} 
 & VASP GPMin & 0.00 & N/A & False & N/A \\ \cline{2-6} 
 & GN-OC20 & 0.00 & N/A & False & N/A \\ \cline{2-6} 
 & GN-OC22 & 79.76 & 67 & True & -0.0001 \\ \cline{2-6} 
 & GNOC-OC20+22 & 0.00 & N/A & False & N/A \\ \hline
\multirow{6}{*}{RbIrO3} & VASP BFGS & 100.00 & 90 & True & 0.0000 \\ \cline{2-6} 
 & VASP CG & 131.11 & 118 & True & -0.0242 \\ \cline{2-6} 
 & VASP GPMin & 71.11 & 64 & True & -0.0012 \\ \cline{2-6} 
 & GN-OC20 & 0.00 & N/A & False & N/A \\ \cline{2-6} 
 & GN-OC22 & 66.67 & 60 & True & -0.0001 \\ \cline{2-6} 
 & GNOC-OC20+22 & 53.33 & 48 & True & -0.0002 \\ \hline
\multirow{6}{*}{Zn2Co3O8} & VASP BFGS & 100.00 & 76 & True & 0.0000 \\ \cline{2-6} 
 & VASP CG & 134.21 & 102 & True & 0.0193 \\ \cline{2-6} 
 & VASP GPMin & 138.16 & 105 & True & -0.0024 \\ \cline{2-6} 
 & GN-OC20 & 0.00 & N/A & False & N/A \\ \cline{2-6} 
 & GN-OC22 & 44.74 & 34 & True & -0.0076 \\ \cline{2-6} 
 & GNOC-OC20+22 & 14.47 & 11 & True & -0.0080 \\ \hline
\end{tabular}
}
\end{table}
\FloatBarrier

\begin{footnotesize}
\begin{longtable}{|c|c|c|c|c|c|}
\caption{VASP and Finetuna results for 3D structures.}
\label{tab:zeolite_summary}\\
\hline
System name &
  Algorithm &
  \begin{tabular}[c]{@{}c@{}}\% DFT\\calls\end{tabular} &
  \begin{tabular}[c]{@{}c@{}}No. DFT\\calls\end{tabular} &
  Converged &
  \begin{tabular}[c]{@{}c@{}}Relaxed energy\\ difference (eV)\end{tabular} \\ \hline
\endfirsthead
\multicolumn{6}{c}%
{{\bfseries Table \thetable\ continued from previous page}} \\
\hline
System name &
  Algorithm &
  \begin{tabular}[c]{@{}c@{}}\% DFT\\calls\end{tabular} &
  \begin{tabular}[c]{@{}c@{}}No. DFT\\calls\end{tabular} &
  Converged &
  \begin{tabular}[c]{@{}c@{}}Relaxed energy\\ difference (eV)\end{tabular} \\ \hline
\endhead
\multirow{2}{*}{\begin{tabular}[c]{@{}c@{}}Cu\_zeolites\_\\ CHA\_T1\_T1\_72\_80\end{tabular}}     & VASP BFGS & 100.00  & 52   & True  & 0.0000  \\ \cline{2-6} 
                                                                                                  & GN-OC20   & 196.15  & 102  & True  & -0.0005 \\ \hline
\multirow{2}{*}{\begin{tabular}[c]{@{}c@{}}Cu\_zeolites\_\\ MFI\_T11\_T11\_272\_274\end{tabular}} & VASP BFGS & 100.00  & 2    & True  & 0.0000  \\ \cline{2-6} 
                                                                                                  & GN-OC20   & 1950.00 & 39   & True  & -0.0015 \\ \hline
\multirow{2}{*}{\begin{tabular}[c]{@{}c@{}}Cu\_zeolites\_\\ MFI\_T2\_T2\_200\_202\end{tabular}}   & VASP BFGS & 100.00  & 12   & True  & 0.0000  \\ \cline{2-6} 
                                                                                                  & GN-OC20   & 691.67  & 83   & True  & -0.0008 \\ \hline
\multirow{2}{*}{\begin{tabular}[c]{@{}c@{}}Cu\_zeolites\_\\ MOR\_T1\_T4\_96\_141\_1\end{tabular}} &
  VASP BFGS &
  100.00 &
  414 &
  True &
  0.0000 \\ \cline{2-6} 
                                                                                                  & GN-OC20   & 12.56   & 52   & True  & -0.0004 \\ \hline
\multirow{2}{*}{\begin{tabular}[c]{@{}c@{}}Cu\_zeolites\_\\ MOR\_T2\_T2\_112\_116\end{tabular}}   & VASP BFGS & 100.00  & 774  & True  & 0.0000  \\ \cline{2-6} 
                                                                                                  & GN-OC20   & 7.62    & 59   & True  & -0.0002 \\ \hline
\multirow{2}{*}{\begin{tabular}[c]{@{}c@{}}Cu\_zeolites\_\\ MOR\_T2\_T4\_112\_141\_1\end{tabular}} &
  VASP BFGS &
  100.00 &
  216 &
  True &
  0.0000 \\ \cline{2-6} 
                                                                                                  & GN-OC20   & 17.13   & 37   & True  & -0.0008 \\ \hline
\multirow{2}{*}{mofs\_zeolites\_000}                                                              & VASP BFGS & 100.00  & 5    & True  & 0.0000  \\ \cline{2-6} 
                                                                                                  & GN-OC20   & 60.00   & 3    & True  & 0.0000  \\ \hline
\multirow{2}{*}{mofs\_zeolites\_001}                                                              & VASP BFGS & 100.00  & 5    & True  & 0.0000  \\ \cline{2-6} 
                                                                                                  & GN-OC20   & 60.00   & 3    & True  & 0.0000  \\ \hline
\multirow{2}{*}{mofs\_zeolites\_003}                                                              & VASP BFGS & 100.00  & 8    & True  & 0.0000  \\ \cline{2-6} 
                                                                                                  & GN-OC20   & 250.00  & 20   & True  & -0.0001 \\ \hline
\multirow{2}{*}{mofs\_zeolites\_004}                                                              & VASP BFGS & 100.00  & 22   & True  & 0.0000  \\ \cline{2-6} 
                                                                                                  & GN-OC20   & 68.18   & 15   & True  & -0.0006 \\ \hline
\multirow{2}{*}{mofs\_zeolites\_005}                                                              & VASP BFGS & 100.00  & 12   & True  & 0.0000  \\ \cline{2-6} 
                                                                                                  & GN-OC20   & 458.33  & 55   & True  & -0.0004 \\ \hline
\multirow{2}{*}{mofs\_zeolites\_007}                                                              & VASP BFGS & 100.00  & 174  & True  & 0.0000  \\ \cline{2-6} 
                                                                                                  & GN-OC20   & 0.00    & N/A  & False & N/A     \\ \hline
\multirow{2}{*}{mofs\_zeolites\_008}                                                              & VASP BFGS & 100.00  & 264  & True  & 0.0000  \\ \cline{2-6} 
                                                                                                  & GN-OC20   & 0.00    & N/A  & False & N/A     \\ \hline
\multirow{2}{*}{zeolites\_np\_00\_AEI\_000}                                                       & VASP BFGS & 100.00  & 399  & True  & 0.0000  \\ \cline{2-6} 
                                                                                                  & GN-OC20   & 21.30   & 85   & True  & -0.0004 \\ \hline
\multirow{2}{*}{zeolites\_np\_00\_AEI\_001}                                                       & VASP BFGS & 100.00  & 336  & True  & 0.0000  \\ \cline{2-6} 
                                                                                                  & GN-OC20   & 33.04   & 111  & True  & -0.0006 \\ \hline
\multirow{2}{*}{zeolites\_np\_00\_AEI\_002}                                                       & VASP BFGS & 100.00  & 457  & True  & 0.0000  \\ \cline{2-6} 
                                                                                                  & GN-OC20   & 40.04   & 183  & True  & -0.0008 \\ \hline
\multirow{2}{*}{zeolites\_np\_00\_AEI\_003}                                                       & VASP BFGS & 100.00  & 385  & True  & 0.0000  \\ \cline{2-6} 
                                                                                                  & GN-OC20   & 32.73   & 126  & True  & -0.0003 \\ \hline
\multirow{2}{*}{zeolites\_np\_00\_AEI\_004}                                                       & VASP BFGS & 100.00  & 401  & True  & 0.0000  \\ \cline{2-6} 
                                                                                                  & GN-OC20   & 14.21   & 57   & True  & -0.0005 \\ \hline
\multirow{2}{*}{zeolites\_np\_00\_AEI\_005}                                                       & VASP BFGS & 100.00  & 267  & True  & 0.0000  \\ \cline{2-6} 
                                                                                                  & GN-OC20   & 34.83   & 93   & True  & -0.0007 \\ \hline
\multirow{2}{*}{zeolites\_np\_00\_AEI\_006}                                                       & VASP BFGS & 100.00  & 297  & True  & 0.0000  \\ \cline{2-6} 
                                                                                                  & GN-OC20   & 20.88   & 62   & True  & -0.0003 \\ \hline
\multirow{2}{*}{zeolites\_np\_00\_AEI\_007}                                                       & VASP BFGS & 100.00  & 403  & True  & 0.0000  \\ \cline{2-6} 
                                                                                                  & GN-OC20   & 24.32   & 98   & True  & -0.0005 \\ \hline
\multirow{2}{*}{zeolites\_np\_00\_AEI\_008}                                                       & VASP BFGS & 100.00  & 276  & True  & 0.0000  \\ \cline{2-6} 
                                                                                                  & GN-OC20   & 18.84   & 52   & True  & -0.0005 \\ \hline
\multirow{2}{*}{zeolites\_np\_00\_AEI\_009}                                                       & VASP BFGS & 100.00  & 436  & True  & 0.0000  \\ \cline{2-6} 
                                                                                                  & GN-OC20   & 0.00    & N/A  & False & N/A     \\ \hline
\multirow{2}{*}{zeolites\_np\_02\_CHA\_000}                                                       & VASP BFGS & 100.00  & 293  & True  & 0.0000  \\ \cline{2-6} 
                                                                                                  & GN-OC20   & 48.46   & 142  & True  & -0.0008 \\ \hline
\multirow{2}{*}{zeolites\_np\_02\_CHA\_002}                                                       & VASP BFGS & 100.00  & 435  & True  & 0.0000  \\ \cline{2-6} 
                                                                                                  & GN-OC20   & 8.05    & 35   & True  & -0.0003 \\ \hline
\multirow{2}{*}{zeolites\_np\_02\_CHA\_003}                                                       & VASP BFGS & 100.00  & 287  & True  & 0.0000  \\ \cline{2-6} 
                                                                                                  & GN-OC20   & 10.80   & 31   & True  & -0.0006 \\ \hline
\multirow{2}{*}{zeolites\_np\_02\_CHA\_004}                                                       & VASP BFGS & 100.00  & 274  & True  & 0.0000  \\ \cline{2-6} 
                                                                                                  & GN-OC20   & 18.61   & 51   & True  & -0.0012 \\ \hline
\multirow{2}{*}{zeolites\_np\_02\_CHA\_005}                                                       & VASP BFGS & 100.00  & 333  & True  & 0.0000  \\ \cline{2-6} 
                                                                                                  & GN-OC20   & 10.51   & 35   & True  & -0.0005 \\ \hline
\multirow{2}{*}{zeolites\_np\_02\_CHA\_006}                                                       & VASP BFGS & 100.00  & 240  & True  & 0.0000  \\ \cline{2-6} 
                                                                                                  & GN-OC20   & 7.50    & 18   & True  & -0.0008 \\ \hline
\multirow{2}{*}{zeolites\_np\_02\_CHA\_007}                                                       & VASP BFGS & 100.00  & 260  & True  & 0.0000  \\ \cline{2-6} 
                                                                                                  & GN-OC20   & 8.46    & 22   & True  & -0.0005 \\ \hline
\multirow{2}{*}{zeolites\_np\_02\_CHA\_008}                                                       & VASP BFGS & 100.00  & 369  & True  & 0.0000  \\ \cline{2-6} 
                                                                                                  & GN-OC20   & 8.67    & 32   & True  & -0.0006 \\ \hline
\multirow{2}{*}{zeolites\_np\_02\_CHA\_009}                                                       & VASP BFGS & 100.00  & 305  & True  & 0.0000  \\ \cline{2-6} 
                                                                                                  & GN-OC20   & 10.49   & 32   & True  & -0.0007 \\ \hline
\multirow{2}{*}{zeolites\_np\_03\_LTA\_000}                                                       & VASP BFGS & 100.00  & 328  & True  & 0.0000  \\ \cline{2-6} 
                                                                                                  & GN-OC20   & 20.12   & 66   & True  & -0.0012 \\ \hline
\multirow{2}{*}{zeolites\_np\_03\_LTA\_001}                                                       & VASP BFGS & 100.00  & 663  & True  & 0.0000  \\ \cline{2-6} 
                                                                                                  & GN-OC20   & 38.31   & 254  & True  & -0.0021 \\ \hline
\multirow{2}{*}{zeolites\_np\_03\_LTA\_002}                                                       & VASP BFGS & 100.00  & 398  & True  & 0.0000  \\ \cline{2-6} 
                                                                                                  & GN-OC20   & 16.33   & 65   & True  & -0.0007 \\ \hline
\multirow{2}{*}{zeolites\_np\_03\_LTA\_004}                                                       & VASP BFGS & 100.00  & 308  & True  & 0.0000  \\ \cline{2-6} 
                                                                                                  & GN-OC20   & 14.94   & 46   & True  & -0.0005 \\ \hline
\multirow{2}{*}{zeolites\_np\_03\_LTA\_005}                                                       & VASP BFGS & 0.00    & N/A  & False & N/A     \\ \cline{2-6} 
                                                                                                  & GN-OC20   & 0.00    & N/A  & False & N/A     \\ \hline
\multirow{2}{*}{zeolites\_np\_03\_LTA\_006}                                                       & VASP BFGS & 100.00  & 303  & True  & 0.0000  \\ \cline{2-6} 
                                                                                                  & GN-OC20   & 18.81   & 57   & True  & -0.0004 \\ \hline
\multirow{2}{*}{zeolites\_np\_03\_LTA\_007}                                                       & VASP BFGS & 100.00  & 341  & True  & 0.0000  \\ \cline{2-6} 
                                                                                                  & GN-OC20   & 51.61   & 176  & True  & -0.0050 \\ \hline
\multirow{2}{*}{zeolites\_np\_03\_LTA\_009}                                                       & VASP BFGS & 100.00  & 500  & True  & 0.0000  \\ \cline{2-6} 
                                                                                                  & GN-OC20   & 12.20   & 61   & True  & -0.0009 \\ \hline
\multirow{2}{*}{zeolites\_np\_04\_MAZ\_001}                                                       & VASP BFGS & 100.00  & 190  & True  & 0.0000  \\ \cline{2-6} 
                                                                                                  & GN-OC20   & 42.63   & 81   & True  & -0.0037 \\ \hline
\multirow{2}{*}{zeolites\_np\_04\_MAZ\_003}                                                       & VASP BFGS & 100.00  & 278  & True  & 0.0000  \\ \cline{2-6} 
                                                                                                  & GN-OC20   & 0.00    & N/A  & False & N/A     \\ \hline
\multirow{2}{*}{zeolites\_np\_04\_MAZ\_004}                                                       & VASP BFGS & 100.00  & 247  & True  & 0.0000  \\ \cline{2-6} 
                                                                                                  & GN-OC20   & 14.17   & 35   & True  & -0.0005 \\ \hline
\multirow{2}{*}{zeolites\_np\_04\_MAZ\_005}                                                       & VASP BFGS & 100.00  & 251  & True  & 0.0000  \\ \cline{2-6} 
                                                                                                  & GN-OC20   & 29.08   & 73   & True  & -0.0034 \\ \hline
\multirow{2}{*}{zeolites\_np\_04\_MAZ\_009}                                                       & VASP BFGS & 100.00  & 2131 & True  & 0.0000  \\ \cline{2-6} 
                                                                                                  & GN-OC20   & 2.86    & 61   & True  & -0.0003 \\ \hline
\multirow{2}{*}{zeolites\_np\_09\_SOD\_000}                                                       & VASP BFGS & 100.00  & 160  & True  & 0.0000  \\ \cline{2-6} 
                                                                                                  & GN-OC20   & 9.38    & 15   & True  & -0.0003 \\ \hline
\multirow{2}{*}{zeolites\_np\_09\_SOD\_001}                                                       & VASP BFGS & 100.00  & 158  & True  & 0.0000  \\ \cline{2-6} 
                                                                                                  & GN-OC20   & 6.96    & 11   & True  & -0.0005 \\ \hline
\multirow{2}{*}{zeolites\_np\_09\_SOD\_002}                                                       & VASP BFGS & 100.00  & 149  & True  & 0.0000  \\ \cline{2-6} 
                                                                                                  & GN-OC20   & 7.38    & 11   & True  & -0.0006 \\ \hline
\multirow{2}{*}{zeolites\_np\_09\_SOD\_003}                                                       & VASP BFGS & 100.00  & 260  & True  & 0.0000  \\ \cline{2-6} 
                                                                                                  & GN-OC20   & 4.62    & 12   & True  & -0.0005 \\ \hline
\multirow{2}{*}{zeolites\_np\_09\_SOD\_004}                                                       & VASP BFGS & 100.00  & 207  & True  & 0.0000  \\ \cline{2-6} 
                                                                                                  & GN-OC20   & 12.56   & 26   & True  & -0.0004 \\ \hline
\multirow{2}{*}{zeolites\_np\_09\_SOD\_005}                                                       & VASP BFGS & 100.00  & 165  & True  & 0.0000  \\ \cline{2-6} 
                                                                                                  & GN-OC20   & 8.48    & 14   & True  & -0.0004 \\ \hline
\multirow{2}{*}{zeolites\_np\_09\_SOD\_006}                                                       & VASP BFGS & 100.00  & 167  & True  & 0.0000  \\ \cline{2-6} 
                                                                                                  & GN-OC20   & 8.38    & 14   & True  & -0.0004 \\ \hline
\multirow{2}{*}{zeolites\_np\_09\_SOD\_007}                                                       & VASP BFGS & 100.00  & 166  & True  & 0.0000  \\ \cline{2-6} 
                                                                                                  & GN-OC20   & 7.83    & 13   & True  & -0.0007 \\ \hline
\multirow{2}{*}{zeolites\_np\_09\_SOD\_008}                                                       & VASP BFGS & 100.00  & 151  & True  & 0.0000  \\ \cline{2-6} 
                                                                                                  & GN-OC20   & 14.57   & 22   & True  & -0.0005 \\ \hline
\multirow{2}{*}{zeolites\_np\_09\_SOD\_009}                                                       & VASP BFGS & 100.00  & 148  & True  & 0.0000  \\ \cline{2-6} 
                                                                                                  & GN-OC20   & 7.43    & 11   & True  & -0.0005 \\ \hline
\multirow{2}{*}{M3\_SAPO-34}                                                                      & VASP BFGS & 100.00  & 93   & True  & 0.0000  \\ \cline{2-6} 
                                                                                                  & GN-OC20   & 89.25   & 83   & True  & -0.0002 \\ \hline
\multirow{2}{*}{M3\_SSZ-13}                                                                       & VASP BFGS & 100.00  & 91   & True  & 0.0000  \\ \cline{2-6} 
                                                                                                  & GN-OC20   & 0.00    & N/A  & False & N/A     \\ \hline
\multirow{2}{*}{M3\_ZSM-22}                                                                       & VASP BFGS & 100.00  & 113  & True  & 0.0000  \\ \cline{2-6} 
                                                                                                  & GN-OC20   & 15.04   & 17   & True  & -0.0008 \\ \hline
\multirow{2}{*}{M3\_ZSM-5}                                                                        & VASP BFGS & 100.00  & 123  & True  & 0.0000  \\ \cline{2-6} 
                                                                                                  & GN-OC20   & 60.98   & 75   & True  & -0.0000 \\ \hline
\multirow{2}{*}{M3a\_SAPO-34}                                                                     & VASP BFGS & 100.00  & 111  & True  & 0.0000  \\ \cline{2-6} 
                                                                                                  & GN-OC20   & 35.14   & 39   & True  & -0.0000 \\ \hline
\multirow{2}{*}{M3a\_SSZ-13}                                                                      & VASP BFGS & 100.00  & 205  & True  & 0.0000  \\ \cline{2-6} 
                                                                                                  & GN-OC20   & 13.17   & 27   & True  & 0.0000  \\ \hline
\multirow{2}{*}{M3a\_ZSM-22}                                                                      & VASP BFGS & 100.00  & 124  & True  & 0.0000  \\ \cline{2-6} 
                                                                                                  & GN-OC20   & 15.32   & 19   & True  & -0.0008 \\ \hline
\multirow{2}{*}{M3a\_ZSM-5}                                                                       & VASP BFGS & 100.00  & 128  & True  & 0.0000  \\ \cline{2-6} 
                                                                                                  & GN-OC20   & 13.28   & 17   & True  & -0.0001 \\ \hline
\multirow{2}{*}{M3b\_MOR8}                                                                        & VASP BFGS & 100.00  & 159  & True  & 0.0000  \\ \cline{2-6} 
                                                                                                  & GN-OC20   & 10.69   & 17   & True  & -0.0000 \\ \hline
\multirow{2}{*}{M3b\_SAPO-34}                                                                     & VASP BFGS & 100.00  & 115  & True  & 0.0000  \\ \cline{2-6} 
                                                                                                  & GN-OC20   & 15.65   & 18   & True  & 0.0004  \\ \hline
\multirow{2}{*}{M3b\_SSZ-13}                                                                      & VASP BFGS & 100.00  & 108  & True  & 0.0000  \\ \cline{2-6} 
                                                                                                  & GN-OC20   & 57.41   & 62   & True  & -0.0001 \\ \hline
\multirow{2}{*}{M3b\_ZSM-22}                                                                      & VASP BFGS & 100.00  & 171  & True  & 0.0000  \\ \cline{2-6} 
                                                                                                  & GN-OC20   & 19.30   & 33   & True  & -0.0000 \\ \hline
\multirow{2}{*}{M3b\_ZSM-5}                                                                       & VASP BFGS & 100.00  & 125  & True  & 0.0000  \\ \cline{2-6} 
                                                                                                  & GN-OC20   & 8.80    & 11   & True  & 0.0000  \\ \hline
M3b\_MOR8                                                                                         & VASP BFGS & 100.0   & 159  & True  & 0.0     \\ \hline
M3b\_MOR8                                                                                         & GN-OC20   & 10.69   & 17   & True  & -0.0    \\ \hline
M3b\_SAPO-34                                                                                      & VASP BFGS & 100.0   & 115  & True  & 0.0     \\ \hline
M3b\_SAPO-34                                                                                      & GN-OC20   & 15.65   & 18   & True  & 0.0004  \\ \hline
M3b\_SSZ-13                                                                                       & VASP BFGS & 100.0   & 108  & True  & 0.0     \\ \hline
M3b\_SSZ-13                                                                                       & GN-OC20   & 57.41   & 62   & True  & -0.0001 \\ \hline
M3b\_ZSM-22                                                                                       & VASP BFGS & 100.0   & 171  & True  & 0.0     \\ \hline
M3b\_ZSM-22                                                                                       & GN-OC20   & 19.3    & 33   & True  & -0.0    \\ \hline
M3b\_ZSM-5                                                                                        & VASP BFGS & 100.0   & 125  & True  & 0.0     \\ \hline
M3b\_ZSM-5                                                                                        & GN-OC20   & 8.8     & 11   & True  & 0.0     \\ \hline
\end{longtable}
\end{footnotesize}